\documentclass[a4paper]{jpconf}

\usepackage{amsmath,amssymb,amsthm}
\usepackage{graphicx}

\newcommand{\ffdeg}{\mathrm{Deg}}

\begin{document}
\title{Numerical calculation of 5-loop QED contributions to the electron anomalous magnetic moment (preprint for ACAT-2019 proceedings)}
\author{Sergey Volkov}
\address{$^1$ Skobeltsyn Institute of Nuclear Physics of Lomonosov Moscow State University, Leninskie gory 1(2), GSP-1, 119991 Moscow, Russia}
\address{$^2$ Dzhelepov Laboratory of Nuclear Problems of Joint Institute for Nuclear Research, Joliot-Curie 6, 141980 Dubna, Moscow region, Russia}
\ead{volkoff\underline{ }sergey@mail.ru, sergey.volkov.1811@gmail.com}
\begin{abstract}
\emph{Preliminary} numerical results of calculating the 5-loop QED contributions to the electron (and muon) anomalous magnetic moment are presented. The results include the total contribution of the 5-loop Feynman diagrams without lepton loops and the contributions of nine gauge-invariant classes that form that set. A discrepancy with known results is revealed. The contributions of the gauge-invariant classes are presented for the first time. The method of the computation is briefly described (with the corresponding references). The calculation is based on the following elements:
\begin{enumerate}
\item a subtraction procedure for removing infrared (IR) and ultraviolet (UV) divergences point by point in Feynman parametric representation before integration; 
\item a nonadaptive Monte Carlo integration method that is founded on probability density functions (PDF) that are constructed for each Feynman diagram individually using its combinatorial structure; 
\item a GPU-based numerical integration with the help of the supercomputer "Govorun" (JINR, Dubna, Russia).
\end{enumerate}
\end{abstract}
\section{Introduction}

The electron anomalous magnetic moment $a_e$ is known with a very high precision. The most accurate measured value~\cite{experiment} is
$$
a_e=0.00115965218073(28).
$$
The most precise theoretical prediction~\cite{kinoshita_atoms} was obtained using the following representation:
$$
a_e=a_e(\text{QED})+a_e(\text{hadronic})+a_e(\text{electroweak}),
$$
$$
a_e(\text{QED})=\sum_{n\geq 1} \left(\frac{\alpha}{\pi}\right)^n
a_e^{2n},
$$
$$
a_e^{2n}=A_1^{(2n)}+A_2^{(2n)}(m_e/m_{\mu})+A_2^{(2n)}(m_e/m_{\tau})+A_3^{(2n)}(m_e/m_{\mu},m_e/m_{\tau}).
$$
The contributions of different parts of this expression were obtained by different researchers. The corresponding value 
$$
a_e=0.001159652181606(229)(11)(12)
$$
based on a relatively independent from $a_e$ measurement of $\alpha$ giving the value
$$
\alpha^{-1}=137.035999046(27);
$$
see ~\cite{alpha_measurement_cs}\footnote{See also the description of another measurement of $\alpha$ in ~\cite{alpha_measurement_rb} that uses rubidium atoms instead of cesium ones in ~\cite{alpha_measurement_cs}.}. The first uncertainty came from the inaccuracy of $\alpha$; the second one came from the numerical uncertainty of $A_1^{(10)}$ due to the statistical error of the Monte Carlo integration; and the last one came from $a_e(\text{hadronic})+a_e(\text{electroweak})$. The first uncertainty is far bigger than the last two ones. However, we should take into account that:
\begin{itemize}
\item These calculations are used for improving $\alpha$; see ~\cite{kinoshita_atoms}.
\item Until recently, $A_1^{(10)}$ had been calculated only by one researcher team ~\cite{kinoshita_atoms}; an independent calculation is required for reliability.
\end{itemize}
Thus, the problem of calculating $A_1^{(2n)}$ is still relevant. We note that the values for $A_1^{(2n)}$, $n=1,2,3,4$ are reliable:
\begin{enumerate}
\item $A_1^{(2)}=0.5$ was first calculated by J. Schwinger in 1948 ~\cite{schwinger1,schwinger2}. A recalculation is contained in many textbooks of quantum field theory.
\item The value $A_1^{(4)}=-0.32847...$ was first obtained by A. Petermann ~\cite{analyt2_p} and independently by C. Sommerfield~\cite{analyt2_z} in 1957. There are some another independent calculations for this value.
\item $A_1^{(6)}$ was being calculated numerically during 1960's and 1970's by different scientific groups ~\cite{levinewright,carrollyao,kinoshita_6}; the most precise value $A_1^{(6)}=1.195\pm 0.026$ for that time was obtained by T.~Kinoshita and P. Cvitanovi\'{c} in 1974. Simultaneously, a work of analytical calculation was in progress. That work was contributed by many different researchers. The final value $A_1^{(6)}=1.181241...$ was obtained in 1996 by E. Remiddi and S. Laporta ~\cite{analyt3}. See also independent calculations in ~\cite{rappl,volkov_2015}.
\item The most precise value $A_1^{(8)}=-1.91298(84)$ obtained by direct numerical integration was presented in 2015 by T. Aoyama, M. Hayakawa, T. Kinoshita and M. Nio ~\cite{kinoshita_10_new}. That value was confirmed and improved by S. Laporta in 2017~\cite{laporta_8} with the help of a semianalytical approach: $A_1^{(8)}=-1.9122457...$. See also independent (but not such precise) calculations in ~\cite{rappl,smirnov_amm} and a calculation for diagrams without lepton loops in ~\cite{volkov_gpu}.
\end{enumerate}

\section{The method}

For reducing the computation time in high-order QED calculations and for making that calculations practically feasible it is very important to remove all IR and UV divergences in Feynman diagrams before integration. It is known that Bogoliubov's R-operation removes all UV divergences in all individual diagrams in Schwinger parametric space ~\cite{hepp,bogolubovparasuk} and in momentum space with noncovariant regularization ~\cite{zimmerman} as well as the Zimmermann forest formula\footnote{This formula was first published in ~\cite{scherbina,zavialovstepanov}. However, the historic name is connected with ~\cite{zimmerman}.} does. The electron anomalous magnetic moment is free from IR divergences, because the on-shell renormalization removes them as well as it removes the UV ones. However, individual diagrams remain IR divergent after applying the direct subtraction on the mass shell. The structure of IR divergences in Feynman parametric space is quite complicated: they can not be recognized by a direct power counting based on a set of Feynman parameters ~\cite{kinoshita_infrared,volkov_2015}. Moreover, sometimes IR and UV divergences can not be separated from each other; see notes in ~\cite{volkov_prd}. For removing both IR and UV divergences different authors developed different subtraction procedures ~\cite{kinoshita_atoms,levinewright,carrollyao,kinoshita_infrared}. We use another one that is based on linear operators that are applied to Feynman amplitudes of UV divergent subdiagrams. A detailed explanation of the procedure is contained in ~\cite{volkov_2015}. Let us recapitulate the advantages of the developed subtraction procedure:
\begin{enumerate}
\item It is fully automated for any order of the perturbation series.
\item It is comparatively easy for realization on computers.
\item It can be represented as a forestlike formula. This formula differs from
Zimmermann's forest formula only in the choice of linear operators and in the way of
combining them.
\item The contribution of each Feynman graph to $A_1^{(2n)}$ can be represented as a single
Feynman parametric integral. The value of $A_1^{(2n)}$ is the sum of
these contributions.
\item Feynman parameters can be used directly, without any
additional tricks.
\end{enumerate}

After applying the subtraction we have a finite integrand for each diagram. Each $n$-loop integrand has $3n-2$ variables\footnote{We use a trick for reducing the number from $3n-1$ to $3n-2$; see ~\cite{volkov_prd}. However, except this trick, we calculate all integrals directly, without any additional reductions.}. However, the shape of these integrands remains unconvenient for integration. The 5-loop integrands have 13 variables; we evaluate these integrals by Monte Carlo. The convergence speed of Monte Carlo integration depends on the choice of the PDF. Adaptive algorithms like VEGAS can fit the PDF to the shape of the integrand. However, these algorithms can adjust only relatively small number of parameters; this can be inefficient for large dimensions. In contrast, for diagrams without lepton loops we propose to use a nonadaptive algorithm that is based on some theoretical considerations about Feynman parametric integrands behavior. The problem of constructing the PDF is very similar to the problem of constructing a good upper bound for the absolute value of the integrand and to the problem of proving the finiteness of renormalized Feynman amplitudes ~\cite{volkov_prd}. Thus, we use some ideas from that area. 

Let $f(z_1,\ldots,z_M)$ be the Feynman parametric integrand. We split all the integration area into the Hepp sectors ~\cite{hepp}:
$$
z_{j_1}\geq z_{j_2}\geq\ldots\geq z_{j_M}.
$$
The PDF is defined as
\begin{equation}\label{eq_pdf}
C\cdot \frac{\prod_{l=2}^M \left(z_{j_l}/z_{j_{l-1}}\right)^{\ffdeg(\{ j_l,j_{l+1},\ldots,j_M\})}}{z_1\cdot z_2\cdot\ldots \cdot z_M},
\end{equation}
where $\ffdeg(s)$ are positive numbers\footnote{Sometimes these numbers are fractional and even less than $1$.} that are defined on all subsets of the set of all internal lines of the diagram except the empty set and the full set. In the hypothetical case when the diagram does not contain UV divergent subdiagrams and we do not take the infrared limit, it can be proved using Speer's lemma ~\cite{speer} that (\ref{eq_pdf}) is an upper bound for $|f(z_1,\ldots,z_M)|$, if $\ffdeg(s)=-\omega(s)$, where $\omega(s)$ is the ultraviolet degree of divergence of the set $s$. When we take the infrared limit, $\ffdeg(s)$ should be decreased. For considering the infrared limit we apply $\omega$ to so-called \emph{I-closures} of sets ~\cite{volkov_prd}. We define the I-closure of a set $s$ as $s\cup s'$, where $s'$ is the set of all photon lines, for which the electron path connecting the ends of this line is contained in $s$. The complete definition of $\ffdeg$ is quite complicated and is not fully justified mathematically; two versions of this definition are presented in ~\cite{volkov_prd,volkov_gpu}. A comparison of the developed algorithm and known adaptive algorithms with respect to the Monte Carlo convergence speed is presented in ~\cite{volkov_prd}.

\section{Preliminary numerical results}

Despite the good convergence, the Monte Carlo integration for the 5-loop case still requires a lot of computer time. For 5-loop QED diagrams without lepton loops this computation was performed using the GPUs NVidia Tesla V100 as a part of the supercomputer "Govorun". That computation led to the preliminary result
\begin{equation}\label{eq_main}
A_1^{(10)}[\text{no lepton loops}]=6.782(113)
\end{equation}
that corresponds to $1\sigma$ limits. For eliminating round-off errors that occur due to numerical subtraction of divergences we used the interval arithmetic with different precisions and speeds; see the details in ~\cite{volkov_prd}. The code for all 3213 integrands with all arithmetics was generated in C++ with CUDA and required 500 GB of disk space in the compiled form. For reliability, the Monte Carlo integration was performed with two different pseudorandom generators from the NVidia CURAND library:
\begin{enumerate}
\item MRG32k3a (Calc 1; 19515 GPU-hours);
\item Philox\underline{ }4x32\underline{ }10 (Calc 2; 6282 GPU-hours).
\end{enumerate}
That calculations gave the results $6.739(132)$ and $6.905(220)$ respectively. The results were statistically combined in (\ref{eq_main}). In total, $1.9\cdot 10^{14}$ Monte Carlo samples were generated and processed. The total calculation time amounted 25797 GPU-hours and the calculation is still in progress. 

Let us remark that there is a significant discrepancy between (\ref{eq_main}) and the result from ~\cite{kinoshita_atoms}:
$$
A_1^{(10)}[\text{no lepton loops: T. Aoyama, T. Kinoshita, M. Nio et al.}]=7.668(159).
$$
Therefore, another independent computation is required.

\begin{figure}[h]
\begin{center}
\includegraphics{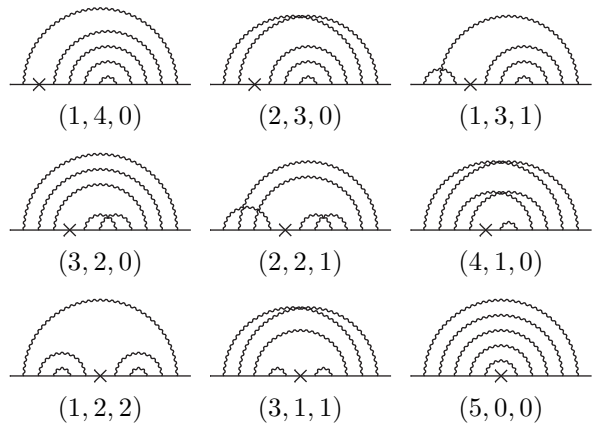}
\end{center}
\caption{Examples of diagrams from the classes $(k,m,n)$.}
\label{fig_examp}
\end{figure}

The developed method allows us to calculate separately the contributions of nine gauge-invariant classes $(k,m,n)$ splitting the whole set. $(k,m,n)$ is the set of all diagrams without lepton loops that have $m$ photon lines to the left from the external photon line, $n$ photon lines to the right (or vice versa), and $k$ photon lines with ends on the opposite sides of the external photon line; see examples in Figure \ref{fig_examp}. Table \ref{table_gauge} contains these preliminary results.
\begin{table}[h]
\caption{Contributions of the gauge-invariant classes $(k,m,n)$ to $A_1^{(10)}{[\text{no lepton loops}]}$.}
\label{table_gauge}
\lineup
\begin{center}
\begin{tabular}{lllllll}
\br
Class & Value$=\sum a_j$ & Calc 1 & Calc 2 & $N_{\text{diag}}$ & $\sum|a_j|$ & $\max|a_j|$\\
\mr 
(1,4,0) & 6.172(42) & 6.158(49) & 6.209(80) & 706 & 1219.7 & 11.8 \\
(2,3,0) & \-0.724(54) & \-0.746(63) & \-0.66(10) & 706 & 3076.8 & 46.2 \\
(1,3,1) & 0.895(43) & 0.854(50) & 1.007(82) & 148 & 3170.3 & 67.5 \\
(3,2,0) & \-0.396(43) & \-0.399(51) & \-0.390(85) & 558 & 2593.5 & 54.9 \\
(2,2,1) & \-2.160(46) & \-2.133(53) & \-2.236(90) & 370 & 3318.0 & 85.8 \\
(4,1,0) & \-1.017(26) & \-1.028(31) & \-0.984(51) & 336 & 1199.3 & 56.7 \\
(1,2,2) & 0.301(25) & 0.312(30) & 0.267(50) & \055 & 1338.4 & 68.7 \\
(3,1,1) & 2.624(30) & 2.628(35) & 2.614(58) & 261 & 1437.2 & 63.5 \\
(5,0,0) & 1.0898(80) & 1.0929(94) & 1.081(15) & \073 & \0137.0 & 19.3 \\
\br
\end{tabular}
\end{center}
\end{table}
Different QED calculations of $a_e$ performed by different researchers showed that relatively small (in absolute value) contributions of gauge-invariant classes are obtained from sums of relatively big contributions of individual Feynman diagrams regardless of the divergence removal method used. The last columns of Table \ref{table_gauge} clearly demonstrate this fact. If $a_j$ is the contribution of a diagram $j$, then the contribution of a class is obtained as $\sum a_j$. For the classes $(k,m,n)$ it is easy to see that $\sum |a_j|$ and $\max |a_j|$ are much greater in absolute value than the contribution. Here $N_{\text{diag}}$ means the quantity of diagrams in the class\footnote{Diagrams that are obtained from each other by changing arrow directions are regarded as one.}.

The developed method also provides a possibility to check the results by parts: the whole set of diagrams can be split into 809 sets for comparison with the direct subtraction on the mass shell in Feynman gauge. The results for these sets will be published in the further papers; see analogous 4-loop results in ~\cite{volkov_gpu} and a comparison of the 2-loop and 3-loop results with the known analytical results in ~\cite{volkov_2015,volkov_gpu}.

\end{document}